# Stability of Glassy and Ferroelectric States in the Relaxors PbMg$_{1/3}$Nb$_{2/3}$O$_3$ and PbMg$_{1/3}$Nb$_{2/3}$O$_3$-12% PbTiO$_3$

Eugene V. Colla, Derek Vigil, John Timmerwilke, and M. B. Weissman

Department of Physics

University of Illinois at Urbana-Champaign

1110 West Green Street, Urbana, IL 61801-3080

D. D. Viehland

Department of Materials Science and Engineering, Virginia Tech, 201 Holden Hall (0237), Blacksburg, Virginia 24061, USA

Brahim Dkhil

Laboratoire Structures, Propriétés et Modélisation des Solides, CNRS-UMR 8580 Ecole Centrale Paris, 92295

Châtenay-Malabry cedex, France

**Abstract:**

The stability of the disordered glassy phase in the relaxors PbMg$_{1/3}$Nb$_{2/3}$O$_3$ and (PbMg$_{1/3}$Nb$_{2/3}$O$_3$)$_{0.88}$(PbTiO$_3$)$_{0.12}$, called PMN and PMN-PT, was investigated by preparing partially polarized samples and allowing them to age at zero field in the temperature range for which the phase is history-dependent. The PMN-PT polarization would spontaneously increase until long-range order formed, first appearing as giant polarization noise. Thus the thermodynamically stable phase in PMN-PT appears to be ferroelectric. In contrast, a PMN sample lacking the sharp first-order field-driven transition found in some other samples spontaneously depolarized, consistent with its glassy state being thermodynamically stable. Detailed thermal depolarization results in PMN show two distinct broad peaks as well as a small fraction of material with a distribution of abrupt melting transitions.

PACS numbers: 77.80.-e, 75.10.Nr, 77.84.-s,

**Introduction**

The perovskite relaxor ferroelectric PbMg$_{1/3}$Nb$_{2/3}$O$_3$ (PMN) and solid solutions of PbTiO$_3$ in it (PMN-PT) have empirical temperature-electric field (T-E) phase diagrams (e.g. Fig. 1) which show a broad hysteretic regime in which either a glassy relaxor state or a somewhat disordered ferroelectric (FE) state [1-5] can exist for long times. In trying to understand the nature of the glassy phase, it would be useful to know whether it is more thermodynamically stable than the ferroelectric phase (as indicated by some theoretical work concerning PMN[6]) or whether the growth of uniformly oriented ferroelectric domains is just inhibited kinetically. For PMN-PT with about 10% PT, this question of relative thermodynamic stability has been raised by measurements which indicate that a ferroelectric phase transition found by surface-sensitive x-ray techniques[7] does not always extend to bulk regions probed by neutron scattering,[8] which leaves open the question of which part is out of equilibrium.

In principle this question might be resolvable (depending on the amount of hysteresis) by accurate measurements of heat capacities and of the latent heat [9] of the first-order field-driven transition to ferroelectric order, but such data are not currently available over the necessary paths in field and temperature. In this paper we report relatively simple time-dependent versions of standard susceptibility and polarization measurements to get at the same thermodynamic information indirectly. We do not here investigate the related question of whether the relaxor regime is separated from the ordinary paraelectric regime by a mere crossover[6] or whether some subtle phase transition (e.g.[10, 11]) may be involved.

The types of measurements we shall use include ones in which we track the spontaneous polarization or depolarization at E=0 of samples whose symmetry has been broken by prior polarization. In effect, these measurements show whether, at E=0, the free energy G increases or decreases as a function of polarization (P) for P below the threshold for conversion to the FE phase. If, throughout the glassy regime at E=0, G decreases as |P| increases, it is safe to conclude that the FE phase rather than the glassy phase is stable. If, on the other hand, G is consistently an increasing function of |P|, the glassy state is more stable. In addition, the patterns of time-dependent polarization and thermal depolarization give some indication of the types of order present, including sizes of domains.

Our current results will show that in a typical PMN-PT (12%) sample the FE phase is thermodynamically more stable than the glassy phase throughout the hysteretic regime. The results on PMN are less complete, but show, in contrast, that in a particular sample in which the field-driven transition is not abrupt, the glassy phase is at least a local free-energy minimum.

**Materials, methods, and characterization**

We compare two samples. One is pure PMN, grown in Rostov-on-Don (Russia) by the Czochralski technique. The other sample is of PMN-12%PT, grown by a modified Bridgman technique by TRS Technologies of State College, PA. This concentration is well below the morphotropic phase boundary, above which all frozen phases are ferroelectric. Both samples were configured as capacitors, with dimensions of thickness ~0.48mm and area ~3mm$^2$ for both samples, oriented with field along a [111] direction. Contacts were made via evaporated Au layers of roughly 100nm thickness on top of Cr layers about 10 nm thick, used to enhance adhesion. The measurement circuitry used externally fixed voltages on the sample and a current-sensing amplifier. Both near-dc polarization current and ac current response to applied ac voltages could be measured simultaneously.

Fig. 2 illustrates the pyroelectric current $I_P$ and the real permittivity $\varepsilon'$ as a function of T on cooling in a field for each sample. The PMN sample showed a fairly sharp crossover to a highly polarized state when cooled in high field, but not the abrupt global first-order forced-ferroelectric (FFE) transition seen in some samples. [1-5] In some other samples[12], this FFE transition is suppressed even more than in this one, presumably by some form of disorder[13]. The absence of an abrupt highly hysteretic form of the transition may provide an advantage for polarization creep measurements, since it allows coexistence of FE-like and glassy regions, allowing the thermodynamically more stable phase to grow without requiring nucleation. This PMN-12%PT sample showed an abrupt FFE transition upon field-cooling.[14] The net polarization was as expected for a simple homogeneous bulk transition, with no indication of any distinct 'skin effect'[8]. The sample was thicker than twice the thickness believed to show special skin-effect behavior previously[8].

**Results**
PMN-PT

As reported previously for other PMN and PMN-PT samples[5], the FFE transition in the PMN-PT sample goes in two steps, but they occur at closely spaced times and very closely spaced temperatures, which need not be resolved for our current purposes. Fig. 3 shows how the transition temperatures depend on the rate of temperature change on cooling and subsequently heating in a field. The freezing temperature shows dramatic rate dependence, detailed in Fig. 4, but the melting temperature shows too little rate dependence to clearly resolve given the limitations of our thermometry. Thus in PMN-PT the melting temperature is at least a candidate for the locus of a true thermodynamic transition, while the empirical freezing temperature is not. The dependence of the freezing temperature on cooling rate very roughly scales with the difference between the melting and cooling temperatures, consistent with some relatively simple kinetic inhibition to forming the frozen state.

Fig. 5 illustrates the aging of $\varepsilon'$ at E=0 in the PMN-PT sample as the sample sits at fixed T after E is set to zero. In large enough |P|, $\varepsilon'$ drops in a fashion similar to that previously observed for samples aging in a large enough E-field to drive a first-order transition to an FE-like state.[3] The sign of the polarization current could be measured for larger starting |P|, showing that the polarization indeed creeps up at E=0. At low initial |P| values, the $\varepsilon'$ aging looks very similar to that found at higher |P|, suggesting but not proving that the same process is at work starting at low |P|.

Although the spontaneous polarization creep is difficult to directly measure (due to background signals) for small initial |P|, the spontaneous increase in polarization can be followed by measuring the integrated depolarization current on subsequent heating to the fully depolarized state. Fig. 6 shows the integrated thermal depolarization in PMN-PT on heating at E=0 as a function of aging time at T=240K and E=0 after field-cooling at 0.17kV/cm. The absolute polarization increases from less than 10% of the saturation polarization to near saturation spontaneously at E=0 on a time scale of about a day. Within the hysteretic regime, we found no P-T point at which the sign of the spontaneous polarization creep at E=0 was opposite to the polarization.

Fig. 7 shows the T-dependent depolarization currents whose integral was shown in Fig. 6. The depolarization current peaks at two different temperatures. The higher peak T is just the standard melting temperature, showing up only in samples aged long enough to partly transform to the FE state, and with magnitude but not temperature dependent on aging time. The lower peak T is strongly dependent on prior aging time, as shown in the inset. The increase of this characteristic temperature under aging closely tracks the increase of |P| shown in Fig. 5, including saturation at long times.

The time dependence of the spontaneous E=0 polarization current, shown in Fig. 8 for the same set of aging runs shown in Figs. 6 and 7, reveals another aspect of the formation of ferroelectric order. Initially, $I_P(t)$ is a smooth function, but after it reaches large values it becomes very noisy, at around 200 s under these conditions. The transition to the noisy state is abrupt and approximately reproducible on the seven runs.

PMN

Our results in PMN concern only the part of the phase diagram below the FFE in a sample for which the FFE is not abrupt. Therefore they are not intended to give definitive results on the thermodynamically stable phase in PMN, but rather to provide a useful contrast to the PMN-PT results and to show some intriguing new qualitative effects.

Fig. 9 illustrates the small spontaneous currents found in similar experiments on the PMN sample. After the initial depolarization transient on setting E=0, there is a small slow creep *reducing* |P|. The sign of this creep at T=180 K was unchanged in the range from |P| = 8 to 12 $\mu C/cm^2$. In this temperature range, some PMN samples undergo the FFE transition at a threshold field of 8.5 $\mu C/cm^2$ with E=3kV/cm. [4].

On subsequent heating, the PMN shows two broad depolarization peaks, centered around 190K and 207 K, as shown in Fig. 10(a). The lower peak is much more strongly reduced by prior aging at E=0. In addition, a small fraction (less than 1%) of the depolarization occurs in abrupt melting steps, appearing as spikes in $I_P(t)$, distributed over a broad range of T. These steps range in size up to about $7 \times 10^{-6} \mu C$-cm dipole moment change, corresponding to the saturation FE polarization in volumes up to about $2 \times 10^{-7} cm^3$, i.e. domains containing over $10^{11}$ polar nanoregions, assuming that the volume of a PNR is no more than $10^{-18} cm^3$, i.e. some $10^4$ unit cells[15]. Fig. 10(b) shows the net polarization change on aging for four hours at T=180 K.

**Discussion**

Some of the interpretation of the PMN-PT results is straightforward. The thermodynamic freezing line is obviously at significantly higher T than the standard empirical line, at least for small E, since the temperature increases significantly as the cooling rate is lowered. Within this hysteretic range even at applied E=0 the polarization spontaneously increases so long as its initial symmetry has been broken. Throughout the hysteretic regime the sample spontaneously creeps toward an FE state. Thus the stable state below the melting line appears to be ferroelectric. The existence of the sharply defined melting temperature is evident even when no fields are applied, since the $\varepsilon(T)$ curves after different aging times all collapse at a well-defined point, near the same T found in the melting of polarized samples. The appearance of a separate freezing line in the empirical phase diagram of PMN-12%PT is thus just a kinetic artifact.

The melting in PMN-PT is nevertheless not entirely understood. A second, lower, characteristic temperature is found in the depolarization current. Unlike the thermodynamic melting point, whose temperature is not history-dependent, the temperature of this broader peak does increase with aging, closely paralleling the total |P|. Exactly what sort of order is involved in this melting stage is not known. Generically, one suspects that this is the melting of some local glassy order, and that if one were to write an effective Hamiltonian for the interactions among the local polarizations (with the global polarization treated as a frozen parameter) the relevant interaction strengths might depend on that average frozen polarization.

The polarization noise during PMN-PT aging shows a dramatic increase above that found in the initial polarization creep long before the average polarization becomes an appreciable fraction of the saturation polarization. The noise increase occurs during a period in which the net polarization rate is nearly constant, indicating that it involves a dramatic increase of the size of the polarizing units, in comparison to the polar nanoregions involved in the initial creep. Individual discrete polarization events involve dipole moments about an order of magnitude larger than those discussed above in the PMN thermal depolarization, again confirming that coherently polarized units far larger than PNR are formed well before the global first-order transition. This polarization noise method may prove useful in other cases where one wants early detection of incipient long-range polar order.

The less complete PMN results may in part have a simple interpretation also. At least in this sample there is no sign that |P| spontaneously increases in the absence of an applied field, regardless of the initial |P|, so long as it is within the glassy phase. Thus in this sample the depolarized glassy state appears to be at least a local free-energy minimum at E=0. In this sample, the sharp depolarization spikes indicate that the partially polarized states are inhomogeneous mixtures of fairly large-scale polarized FE-like regions and glassy regions. Since there are seed nuclei of the FE state, but no tendency to spontaneously polarize further, it appears that the glassy regions are more stable, and thus could represent a true thermodynamic equilibrium phase. However, these data cannot rule out the possibility that in samples which show an abrupt FFE accompanied by an overall lattice distortion the FFE state might be thermodynamically stable even at E=0. The field-cooled version of this FFE state, unlike the one prepared under applied field after zero-field cooling, has not been found to depolarize when the field is set to zero.[16] Whether that is because the field –cooled state in samples with a sharp transition remains the equilibrium state at E=0 or whether it simply lacks nuclei of an equilibrium glassy state remains to be resolved.

The differences between the samples which show a global FFE transition and those that do not are not yet fully understood, although some effects of strain fields with long-range correlations are suspected of inhibiting the transition.[13] [14] The distribution of local melting temperatures may represent a distribution of genuine thermodynamic melting temperatures in the presence of quenched disorder[17, 18], if the PMN-PT results are taken as a guide to the interpretation.

The existence in PMN of two distinct depolarization peaks in $I_P(T)$ with different behavior under aging indicates the presence of two distinct types of short-range order. This result represents the latest in a long string of data on aging and other effects in cubic relaxors (e.g. [5, 19, 20]) which require multiple types of order. The results fit well with results showing spatial inhomogeneity, with a mixture of polar nanoregions regions and regions with some sort of glassy correlations.[11, 21]

The lower-temperature depolarization peaks in PMN and PMN-PT share some features. Both are strongly sensitive to zero-field aging, although in different ways. A more complete set of low-PT samples would be helpful in understanding the relation between these peaks. It is likely, however, the small-scale spinglass-like order is dominant in these low-temperature aging effects.

Overall, our results support earlier claims[7], based on scattering data, that PMN-10% PT spontaneously forms FE order, in contrast to pure PMN. It is possible that the cubic structures found for PMN-10% PT via neutron scattering[8] reflect a non-equilibrated bulk region. It also possible that the slightly higher (12%) PT concentration in our sample was just large enough to cross the equilibrium phase boundary to a ferroelectric state, or that in this PT range differences in quenched randomness between samples are important. It would be interesting to see neutron scattering results on the symmetry of samples aged at E=0 after prior field biasing.

**Acknowledgements:**

This work was funded by NSF DMR 02-40644 and used facilities of the Center for Microanalysis of Materials, University of Illinois, which is partially supported by the U.S. Department of Energy under grant DEFG02-91-ER4543.

**Figures**

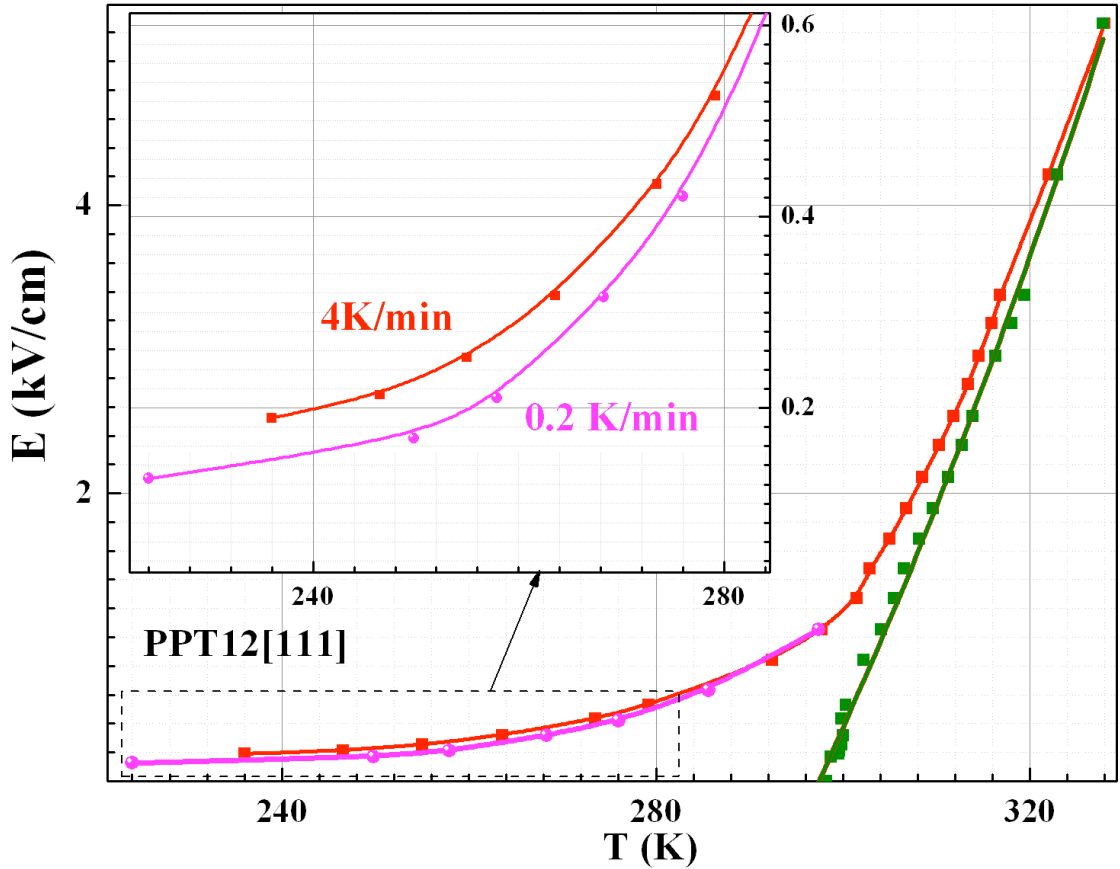

Figure 1. (color online) An empirical E-T phase diagram is shown for the PMN-PT 12% sample described in this paper. The lines on the left are the temperatures at which an abrupt polarization increase occurs upon field cooling. The leftmost one is measured at a cooling rate of 4 K/min, the next one at a rate of 0.2 K/min. The line on the right is the temperature at which an abrupt depolarization occurs. The inset shows the enlarged low-T fragment of the phase diagram.

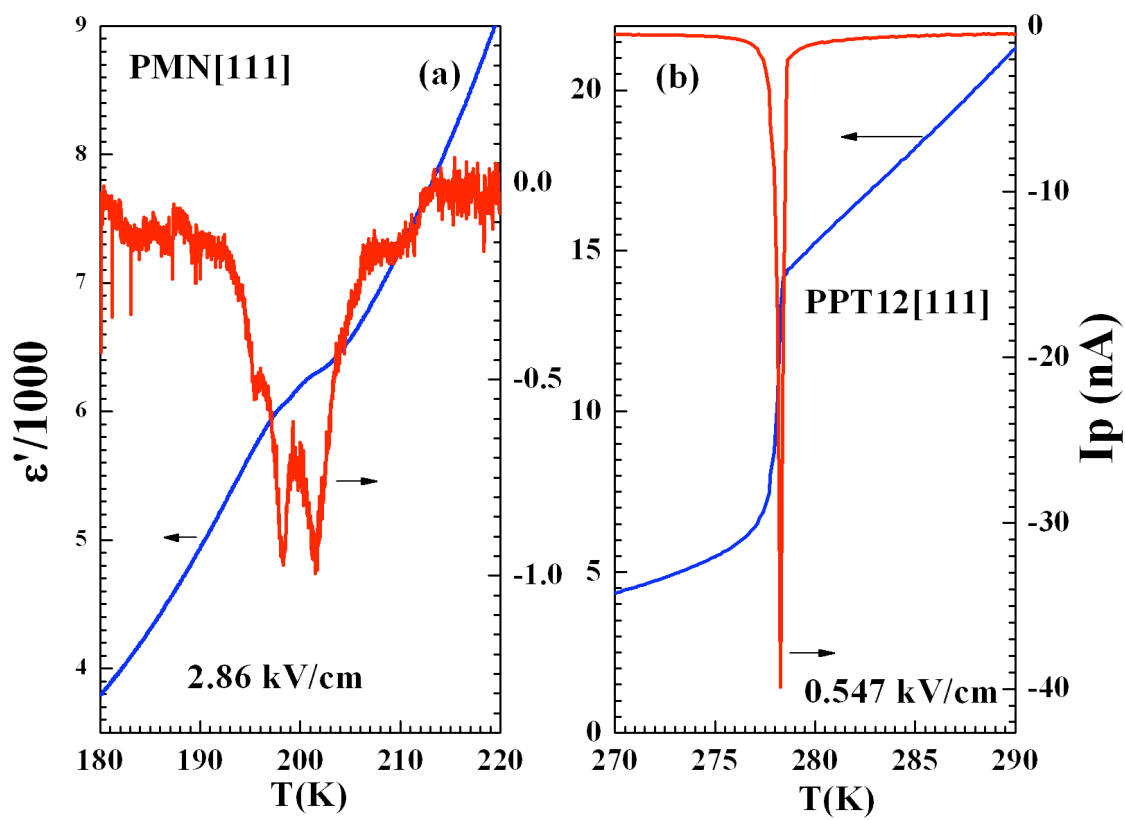

Figure 2. (color online) The real dielectric constant ε' and the polarization current $I_P$ are shown for each sample on cooling at 4 K/min in the fields of 2.86 kV/cm (PMN) and 0.547 kV/cm (PMN-PT).

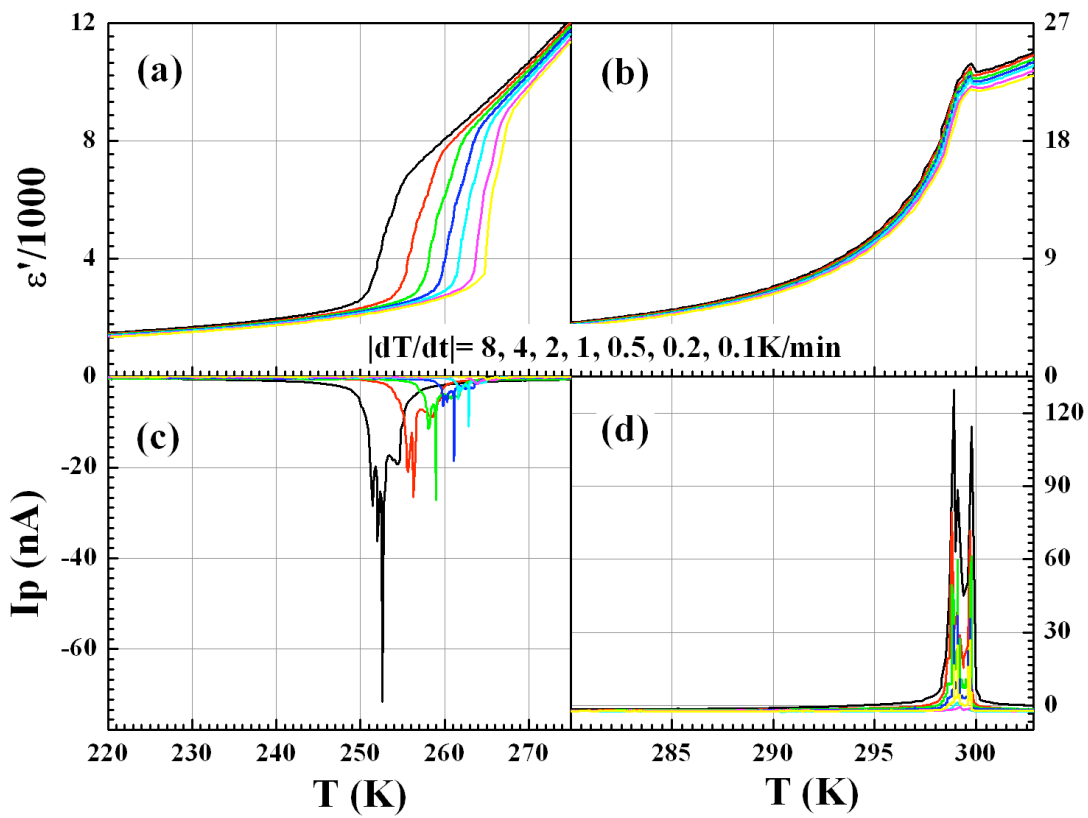

Figure 3 (color online). Portions of ε'(T) (a and b) and $I_P$ (c and d) are shown for PMN-PT on cooling (a and c) and heating (b and d) at a range of different rates in a field of 0.27 kV/cm (the same for cooling and heating).

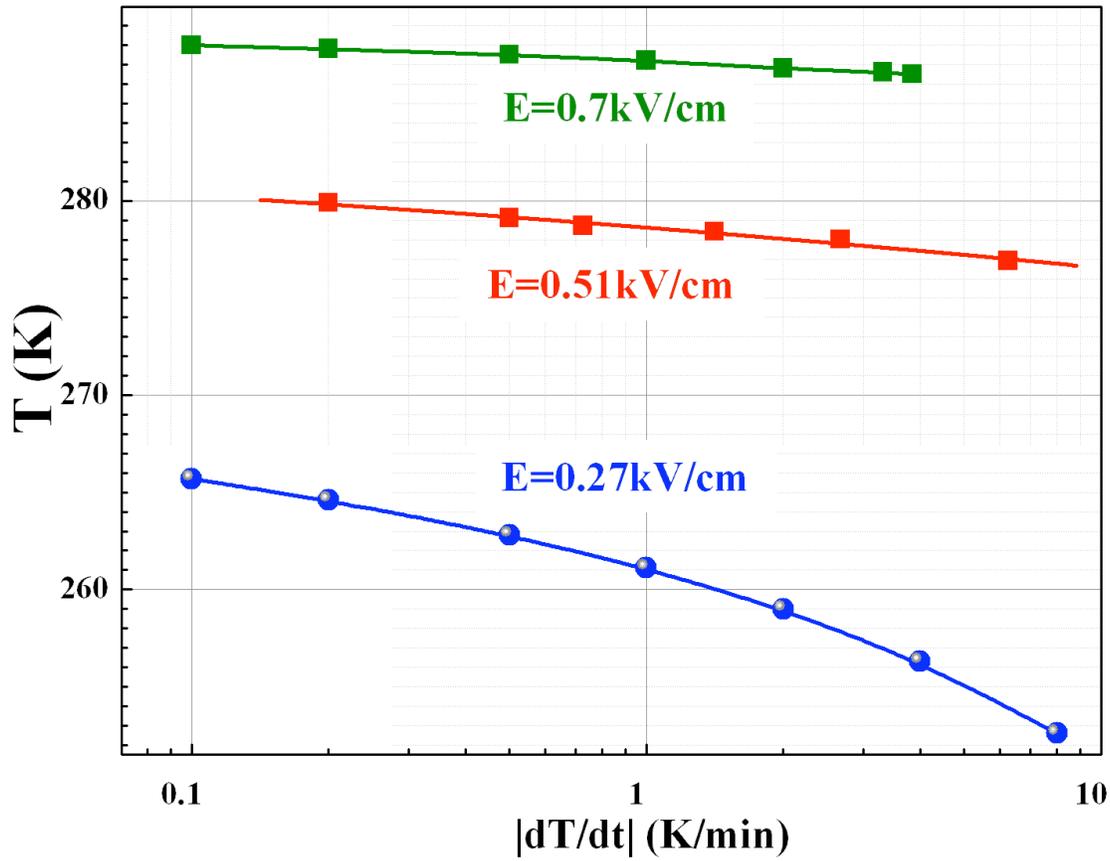

Figure 4. (color online) The cooling-rate dependence of the freezing temperature for PMN-PT is shown at three different fields.

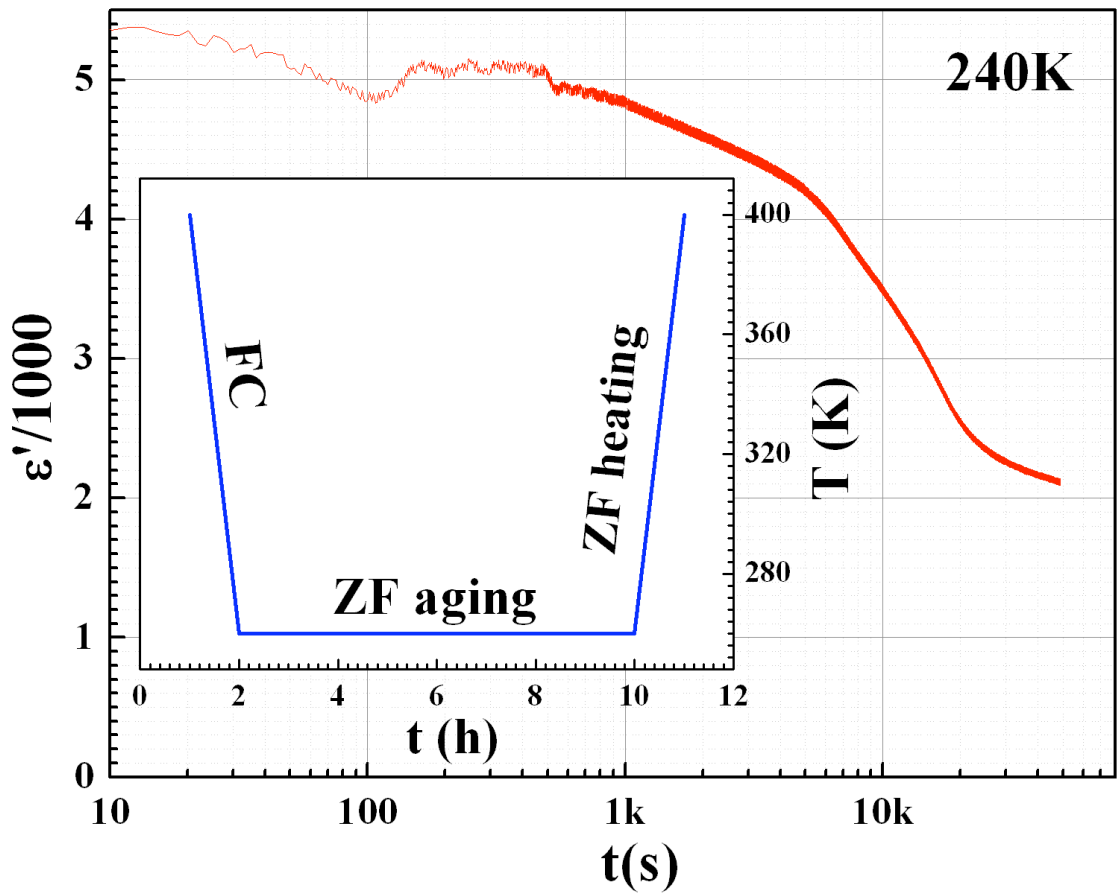

Figure 5. (color online) The inset describes the field-temperature history employed in the data shown here and below. The main figure shows ε' as a function of aging time at E=0 and T=240K after the sample was cooled down in FC regime at 0.11 kV/cm and before subsequent reheating.

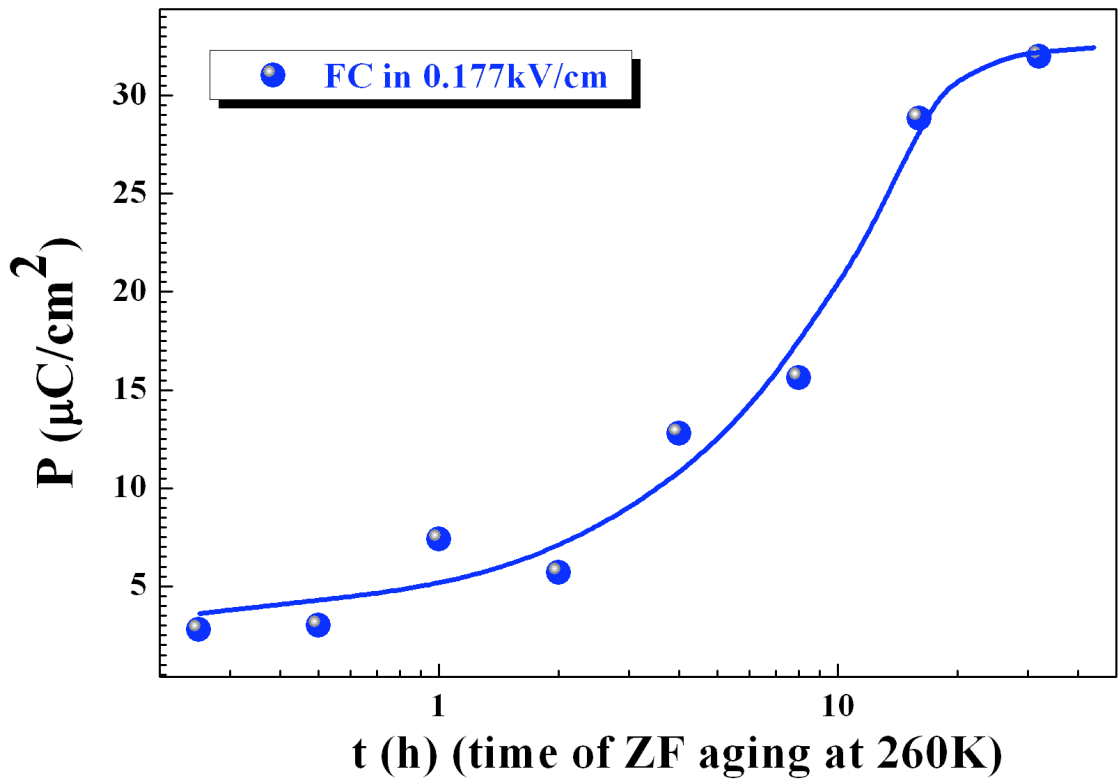

Figure 6. (color online) The dependence of the net polarization (measured from the integrated pyroelectric current on warming) is shown as a function of aging time at T= 260K and E=0 after prior cooling at E=0.177 kV/cm.

Colla 15 3/28/07 11:28 AM

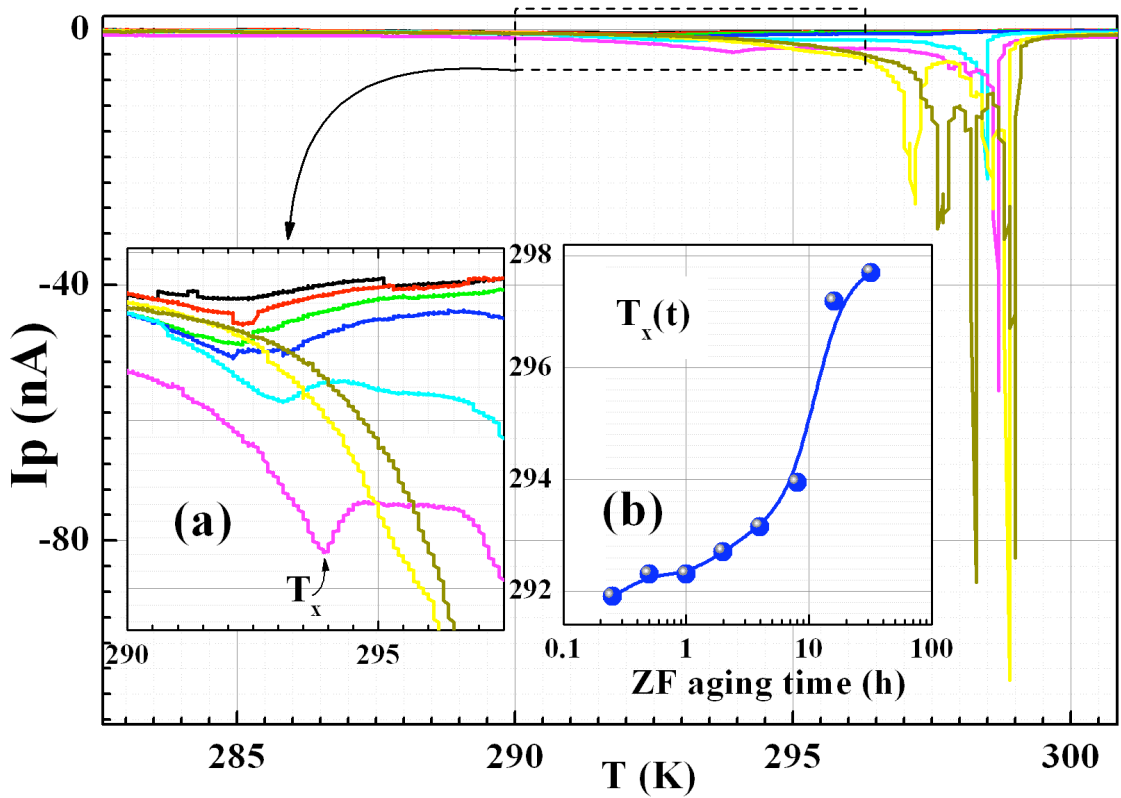

Figure 7. (color online) The pyrocurrent from PMN-PT is shown on warming after different aging times (as in Fig. 6) at E=0 and T=260K after cooling at E= 0.177 kV/cm. Inset (a) is a blow up of the first peak in $I_P(T)$, with increasing peak temperatures, $T_X$, corresponding to longer prior aging. Inset (b) shows how $T_X$ varies with aging time.

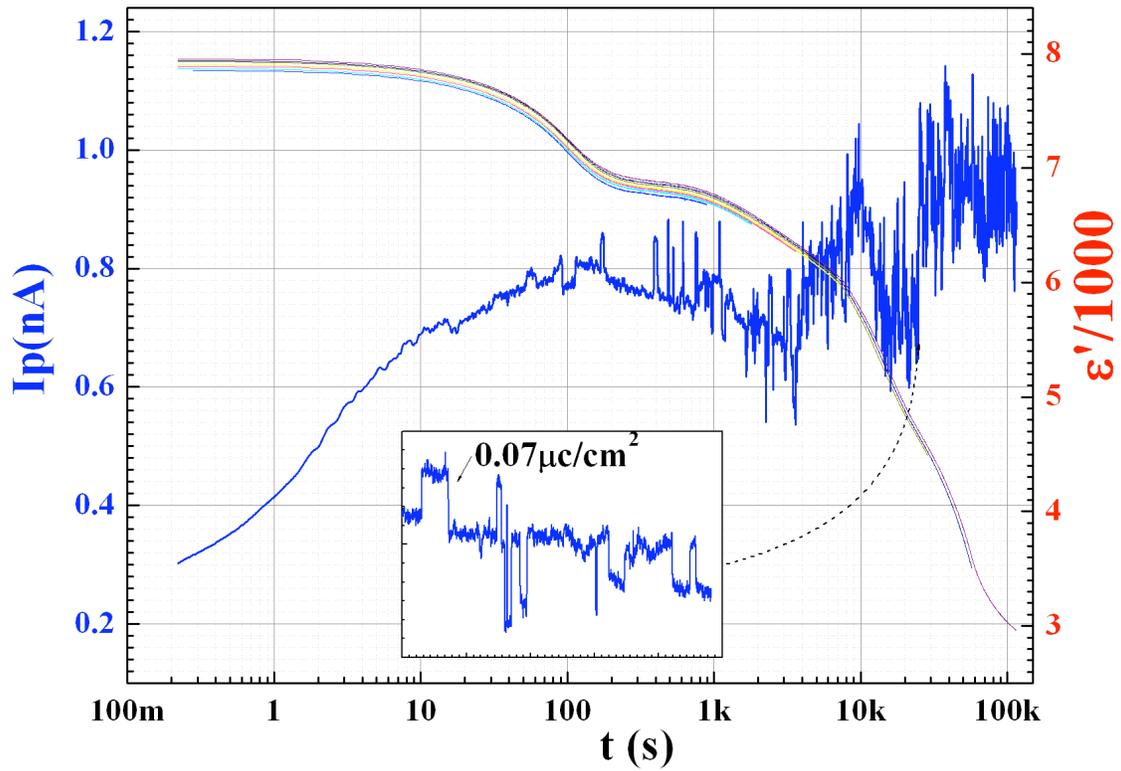

Figure 8. (color online) The dependence of ε' and $I_P$ on aging time at E=0 and T=260K are shown for the PMN-PT after prior cooling at E 017 kV/cm. The blow-up shows an example of the time-dependence of the current noise.

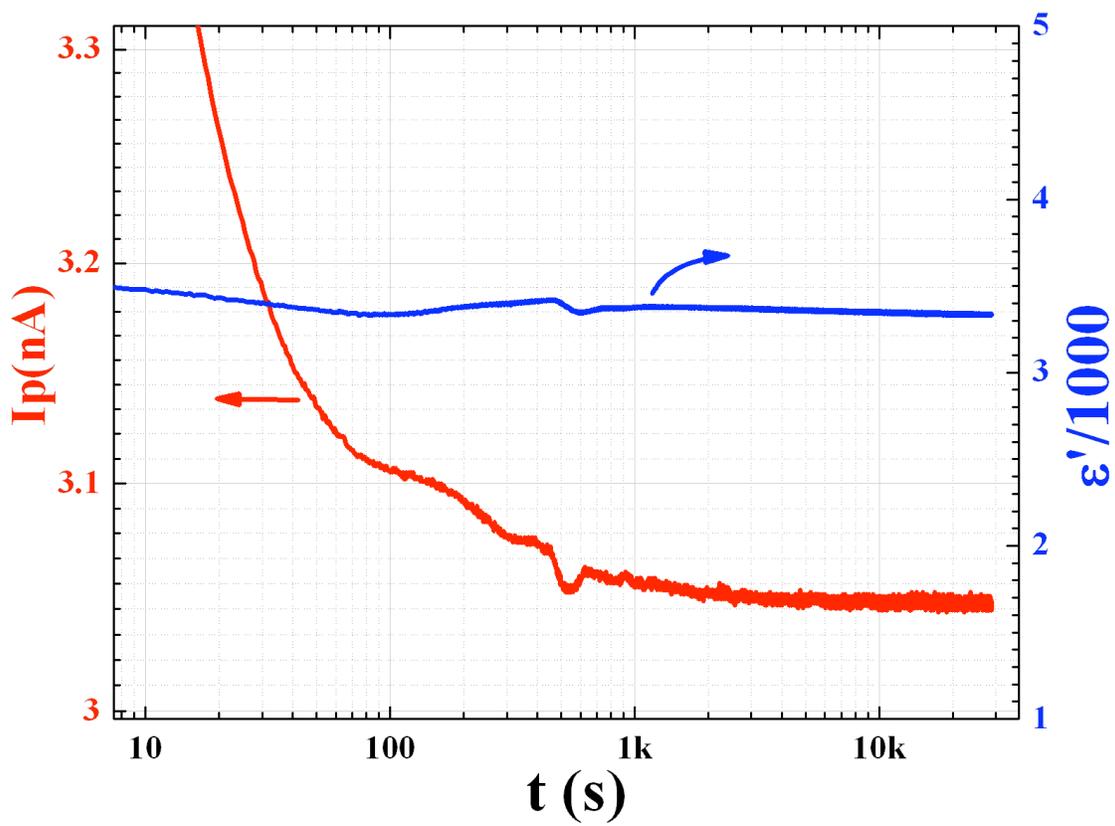

Figure 9. (color online) The time-dependence of ε' and $I_P$ of PMN are shown during aging at E=0 and T=180K after cooling at E=1.63 kV/cm.

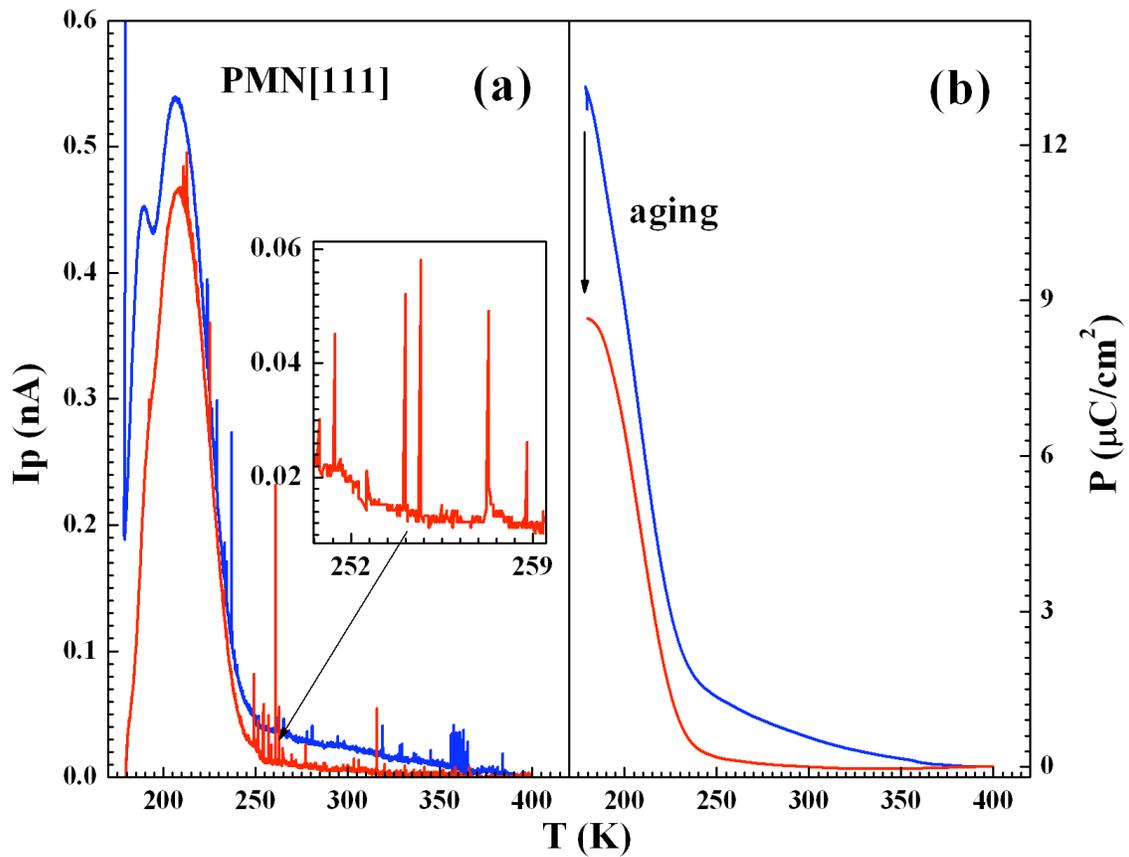

Fig. 10. (color online) Part (a) shows $I_P$ for PMN as a function of T on warming after field-cooling and zero-field aging for a short time (top curve) and four hours (lower curve) after field cooling at E=1.63 kV/cm. The blow-up shows the discrete depolarization current spikes. Part (b) shows the integrated polarization inferred from $I_P$.